\documentclass[a4paper,12pt]{article}
\usepackage{amssymb,amscd,amsmath}
\usepackage{graphicx}

\def\j#1#2#3#4{{\it #1} {\bf #2} #3 #4}

\def\prl{Phys. Rev. Lett.}
\def\pl{Phys. Lett.}
\def\np{Nucl. Phys.}

\def\pr{Phys. Rev.}

\def\ap{Ann. Phys.}

\def\be{\begin{equation}}
\def\ee{\end{equation}}
\def\bea{\begin{eqnarray}}
\def\eea{\end{eqnarray}}

\def\sote{SO(10)}
\def\sufi{SU(5)}
\def\susi{SU(6)}
\def\suth{SU(3)}
\def\suthc{SU(3)_\mathrm{C}}
\def\suthh{SU(3)_\mathrm{H}}
\def\sutw{SU(2)}
\def\sutwl{SU(2)_\mathrm{L}}
\def\uon{U(1)}
\def\uonb{U(1)_\mathrm{B}}
\def\uonc{U(1)_\mathrm{C}}
\def\uony{U(1)_\mathrm{Y}}
\def\ie{\textit{i.e.} }
\def\etal{\textit{et.al.} }

\title{Grand Unified Theory based on the $\susi$ symmetry}

\author{A. Hartanto$^{\mathrm{a}}$\footnote{Email : andreashartanto@cbn.net.id}
\, \, and \, \, 
L.T. Handoko$^{\mathrm{a,b}}$\footnote{Email : handoko@lipi.go.id, handoko@fisika.lipi.go.id, handoko@fisika.ui.ac.id}}
\date{}

\begin{document}

\maketitle

\begin{picture}(0,0)
       \put(310,160){FISIKALIPI-04007}
       \put(310,140){FIS-UI-TH-05-02}
\end{picture}

\thispagestyle{empty}

\begin{center}
\begin{small}
\noindent
$^\mathrm{a)}$ Department of Physics, University of Indonesia\footnote{http://www.fisika.ui.ac.id}, Kampus UI Depok, Depok 16424, Indonesia \\
\vspace*{2mm} 
$^\mathrm{b)}$ Group for Theoretical and Computational Physics, Research Center for Physics, Indonesian Institute of Sciences\footnote{http://www.fisika.lipi.go.id}, Kompleks Puspiptek Serpong, Tangerang 15310, Indonesia \\
\end{small}
\end{center}

\vspace*{5mm}

\begin{abstract}
We present a complete set of generators for the rank 5 special 
unitary group, $\susi$, to unify strong, weak and electromagnetic 
interactions. The unification is realized through the breaking pattern 
of $\susi \rightarrow \suthc \otimes \suthh \otimes \uonc$ followed by 
$\suthh \rightarrow \sutwl \otimes \uonb$. All known elementary particles 
and its quantum numbers are well accomodated in its $\{6\}$ and $\{15\}$ 
multiplets. These multiplets require a new neutral fermion which should be 
assigned as the heavy Majorana neutrino to realize the 
seesaw mechanism naturally in the minimal scenario of this model. 
\end{abstract}

\vspace*{5mm}
\noindent
PACS : 12.10.Dm, 12.60.Cn, 11.30.2j, 02.20.Qs \\

\newpage

\section{Introduction}
\label{sec:intro}

Nowadays, all phenomenons in the high energy physics have been explained 
within the standard model (SM) which is a gauge theory based on the 
$\suthc \otimes \sutwl \otimes \uony$ symmetry \cite{gws}. 
This set of symmetry represents strong, weak and electromagnetic 
interactions in a universal framework. In contrast to the weak and 
electromagnetic interactions which have been successfully unified in 
the electroweak theory based on $\sutwl \otimes \uony$ symmetry, the 
strong interaction with $\suthc$ symmetry remains independent 
from the others. 

So far, the electroweak theory is in impressive agreement with the 
most of experimental 
observables \cite{pdg}. However, there are recently several experimental 
results which disagree with the SM's predictions, 
as the oscilation in the neutrino sector \cite{massiveneutrino} and 
the discrepancy in the NuTeV measurement \cite{nutev}. 
There are also undergoing or 
forthcoming experiments to measure the double $\beta$ decay \cite{betadecay}, 
to search the Higgs particle(s) needed to break the symmetry \cite{lhc}, 
to relate the high energy phenomenon with the cosmology one  
and so on. All of them have been expected to be able to distuingish 
some physics beyond the SM. 
As mentioned above, the SM is lacking of explaining  
the unification of three gauge couplings at a particular scale, especially 
under an assumption that our nature should be explained by a 
single unified theory, the so called grand unified theory (GUT). 

In order to realize GUT at some scale, most of works in the last decades 
have dealed with gauge theory inspired by the successfull 
electroweak theory. Those theories assumed the gauge invariance  
under particular symmetries larger than the SM's one, but contain 
$\suthc \otimes \sutwl \otimes \uony$ as a part of its subgroups at 
electroweak scale. Just to mention, the most famous models 
in this category are $\sufi$ \cite{su5} and $\sote$ \cite{so10}. 
Thereafter there are a lot of variants of these models including the 
supersymmetrized ones. There were also few works on $\susi$ GUT with 
and without supersymmetry  \cite{su6, su6nn}. 

In this paper, we follow the same approach to extend the SM by 
introducing larger symmetry based on $\susi$ group. Since this 
group has rank 5, same as $\sote$, it will provide a new alternative to 
realize gauge unification beside $\sote$ GUT. Higher rank than $\sufi$  
implies that the $\susi$ GUT has several scales before breaking down to 
$\suthc \otimes \sutwl \otimes \uony$. This is an important feature to 
avoid too fast proton decay \cite{andreas1}. 

Before going further, we would like to point out the differences  
between the present paper and the previous works in $\susi$ GUT 
\cite{su6, su6nn}.
The pioneering work \cite{su6} has dealed with the $\susi$ GUT which 
breaks down to the SM through the breaking pattern in the first stage as 
$\susi \rightarrow \sufi \otimes \uon$ by introducing an extra 
$\uon$ gauge symmetry.  
This was highly motivated by a desire to incorporate the $\sufi$ GUT 
which was the most potential candidate for GUT that time. 
Some subsequent works \cite{su6nn} then extended the model to its 
supersymmetric versions. 
In contrast, in the present work we follow different breaking pattern, 
\ie $\susi \rightarrow \suth \otimes \suth \otimes \uon$. 
Moreover, we also begin 
from constructing the $\susi$ group itself, that is determining 
explicitly the generators which could realize the assumed symmetry 
breaking. Using the extended Gell-Mann Okubo relation which is derived 
from the Cartan sub-algebra in the group, we assign the fermions in 
the appropriate multiplets. Actually,  we propose completely different fermion 
multiplets in the $\{6\}$ and $\{\overline{15}\}$ representations. 
Therefore, the model is entirely different from the previous $\susi$ GUT 
models. 

The paper is organized as follow. First, we discuss appropriate 
symmetry breakings in $\susi$ GUT in Sec. \ref{sec:ssb}.  
Based on these breaking patterns, the basic of $\susi$ group and its 
generators are then given in Sec. \ref{sec:su6}. 
Before presenting the extended Gell-Mann Okubo relation, we perform a 
detail study of the quantum numbers contained in the model in 
Sec. \ref{sec:quantum}. Using this extended Gell-Mann Okubo relation, 
we propose a new particle assignment in the $\susi$ multiplets, \ie 
$\{6\}$  and $\{\overline{15}\}$ in Sec. \ref{sec:particle}. We show that 
in this configuration at least a new neutral fermion is naturally required. 
This could then be interpreted as the heavy Majorana neutrino that is 
crucial in the so-called seesaw mechanism needed to explain very 
light observed neutrinos.

\section{Pattern of symmetry breaking}
\label{sec:ssb}

First of all, determining the pattern of symmetry breaking in a 
GUT model is a crucial step. 
In the case of $\susi$ group, concerning only 
the sub-matrices of its generators, intuitively there are several 
possibilities to break the symmetry, for example
\be
  \susi \rightarrow \left\{
  \begin{array}{l}
   \sufi \otimes \uon \\
   \sutw \otimes \sutw \otimes \sutw \otimes \uon \otimes \uon \\
   \suth \otimes \suth \otimes \uon
  \end{array}
  \right.  \; .
  \label{eq:ssb}
\ee
The first choice is clearly similar to the known $\sufi$ GUT where 
it is followed by the breaking pattern of 
$\sufi \rightarrow \suth \otimes \sutw \otimes \uon$ 
to obtain the SM. This breaking 
pattern has been introduced by \cite{su6}, 
however this is not much preferred due to too fast proton decay. 
On the other hand, the second example can be excluded since it is not 
able to accomodate the SM. Then the last one is the only pattern we should 
choice and it has actually not been studied so far. 

At this present stage, we can straightforward put the first $\suth$ as  
$\suthc$ representing the strong interaction, while the second one 
should break further to $\sutwl \otimes \uony$ to reproduce the 
electroweak theory. So, there are two stages to break $\susi$ down to 
the electroweak scale, 
\bea
   \susi & \rightarrow & \suthc \otimes \suthh \otimes \uonc \nonumber \\
         & \rightarrow & \suthc \otimes \sutwl \otimes \uonb \otimes \uonc \; ,
   \label{eq:ssbsu6}
\eea
where $H$ denotes a new quantum number which we later on call as 
hyper-isospin. The combination of quantum numbers induced by 
$\uonb$ and $\uonc$ will 
reproduce the familiar hypercharge associated with $\uony$ in the electroweak 
theory. These points will be clarified in detail in Sec. \ref{sec:quantum}.

Next, we should consider the fundamental representations and the minimal 
multiplets to accomodate the particle contents. 
The fundamental representations of $\susi$ group is represented as 
$\{6\}$ and its anti-symmetric $\{\overline{6}\}$. A tensor product of 
two fundamental representations gives, 
$\{6\} \otimes \{6\} = \{21\} \oplus \{\overline{15}\}$. Following the 
general requirement for the anomaly free combination of representations 
of fermions in any particular $SU(N)$ group \cite{anofree}, one should 
choose the combination of $2 \{6\} \oplus \{\overline{15}\}$ in the case of 
$\susi$. The second $\{6\}-$dimensional representation comes up from 
the decomposition of $\{21\}$ in the above tensor product.
Therefore we can conclude here that the fermions must be assigned in 
these multiplets, namely sextet ($\{6\}$) and decapentuplet ($\{15\}$). 
The particle contents in each multiplet will be given in Sec. 
\ref{sec:particle} after deriving the extended Gell-Mann Okubo relation 
in Sec. \ref{sec:quantum}. 

\section{$\susi$ group}
\label{sec:su6}

In this section, we construct the generators for $\susi$ group. 
In general, the generators for $SU(N)$ group can be determined  
using the existing generators of $SU(N-1)$ group and expanding  
its $(N-1) \times (N-1)$ matrices to be $N \times N$ matrices \cite{stancu}. 
Then there are three considerable types of matrices which could 
form an $SU(N)$ group, 
\be
  \lambda_i = \left\{
   \begin{array}{lcl}
   \left(\begin{array}{ccccc|c}
           &  &  &   &                                  & 0 \\
              &  &  &   &         &  \\
          \multicolumn{5}{c|}{\tilde{\lambda}_i}   & \vdots\\
           &  &  &   &                                  & \\
           &  &  &   &                                  & 0\\ 
           \hline
           0 & & \cdots &  &         0                  & 0 \\
          \end{array} 
   \right) & , \; \mathrm{for} & i = 1, 2, \cdots, (N-1)^2 - 1 \\
     & & \\
   \left(\begin{array}{ccccc|c}
           &  &  &   &                                  & 0\\
                     &  &  &   &               & \vdots\\
          \multicolumn{5}{c|}{(0)_{(N-1)\times (N-1)}}        & a_{jN}\\
           &  &  &   &                                  & \vdots\\
           &  &  &   &                                  & 0\\ 
           \hline
          \multicolumn{5}{c|}{0 \cdots a_{Nj} \; \cdots \; 0}  & 0\\
          \end{array} 
   \right) & , \; \mathrm{for} & (N-1)^2 - 1 < i < N^2 - 1 \\
     & & \\
    \lambda_{N^2-1} & , \; \mathrm{for} & i = N^2 - 1
   \end{array}
  \right. \; , 
  \label{eq:sun}
\ee
where $\tilde{\lambda}_i$ is the $i-$th generator of $SU(N-1)$ group and 
$a_{jN} = a^\ast_{Nj} = 1$ or $-i$ with $j = 1,2, \cdots, N-1$. 
This confirms that the total number of generators in an $SU(N)$ group 
equals to 
$\left[ (N-1)^2 - 1 \right] + \left[ 2 \times (N-1) \right] + 1 = N^2-1$.
Note that special (physical) 
consideration must be taken to determine the last generator, \ie  
$\lambda_{N^2-1}$ beside the basic mathematical requirement 
$\mathrm{tr} (\lambda_i \lambda_j) = 2 \delta_{ij}$. 
Of course one should remark that the 
order of numbering the generators can be changed for the sake 
of convenience due to some physical considerations as discussed soon. 
Now we are ready to move forward to the case of $\susi$ group. 

Throughout the paper we use the notation $\overline{\lambda}_i$ to indicate 
the generators of $\suth$ (Gell-Mann matrices \cite{gellmann}), 
$\tilde{\lambda}_i$ for $\sufi$, 
$\lambda_i$ for the $\susi$ and $\sigma_{1,2,3}$ for the Pauli matrices. 
We start from the well-known generators of $\sufi$ \cite{mohapatra}. 
It is considerable to bring the $\tilde{\lambda}_{1,\cdots,20}$ as 
they are and extend them to be $\lambda_{1,\cdots,20}$ by adding the 
$6-$th rows and columns with null elements. This implies that the 
color quantum number is preserved as the conventional quantum chromodynamics 
(QCD), \ie the upper left $3 \times 3$ block still 
represents the $\suthc$ symmetry. Since the last generator should 
form the Cartan sub-algebra which determines the (physically meaningfull) 
eigen values, that is having non-zero diagonal elements, we eliminated the 
$\tilde{\lambda}_{24}$. Instead of that we put $\lambda_{21,\cdots,26}$ 
as the type of extended $\sigma_1$ and $\sigma_2$ matrices filling in 
the upper-right and lower-left $3 \times 3$ blocks.  

Further, $\tilde{\lambda}_{21,22,23}$ are kept and extended to be 
$\lambda_{27,28,29}$ to represent the $\suthh$ group after the first step of 
symmetry breaking in Eq. (\ref{eq:ssbsu6}). The extended $\sigma_1$ and 
$\sigma_2$ types with its 
non-zero elements filling the last rows and columns in the lower-right 
$3 \times 3$ block form $\lambda_{30,31,32,33}$.

Since $\susi$ is a rank 5 group, it should have five generators form 
its Cartan sub-algebra. In a more technical term, there must be five generators 
with non-zero diagonal elements. Since we already have three of them 
($\lambda_{3,8,29}$), therefore we should define the remaining two diagonal 
generators. From the fact that $\lambda_{27,\cdots,33}$ have the same form 
as the extended $\overline{\lambda}_{1,\cdots,7}$ of $\suth$, it is 
then appropriate to choose $\lambda_{34}$ as the extended form of 
$\overline{\lambda}_8$. 

As mentioned earlier, we must take physical considerations to 
determine the remaining $\lambda_{35}$. Concerning the first step of 
symmetry breaking in Eq. (\ref{eq:ssbsu6}), $\lambda_{35}$ should 
reflect the quantum number of $\uonc$ and be independent from 
both $\suthc$ and $\suthh$. It yields that, 
\be
   \lambda_{35} = \frac{1}{\sqrt{3}} \left(
   \begin{array}{c|c}
           (-1)_{3 \times 3} & (0)_{3 \times 3}\\ 
          \hline
          (0)_{3 \times 3}  & (1)_{3 \times 3}
           \end{array} \right) \; .
   \label{eq:l35}
\ee

Finally, the generators for $\susi$ group can be defined in a common way 
using these matrices as follows, 
\be
  F_i \equiv \frac{1}{2} \, \lambda_i \; \; \; \; (i : 1, \cdots, 35) \;, 
\ee
which satisfies the relation $[F_i, F_j] = i f_{ijk} F_k$ 
with $f_{ijk}$ is the structure constant respectively. 
Complete expressions for all matrices are given in the Appendix.

Before going on to the next section, we would like to make several remarks 
here, 
\begin{itemize}
\item The Gell-Mann like matrices $\lambda_{27,\cdots,34}$ with non-zero 
elements in the lower-right $3 \times 3$ block represents the $\suthh$ 
at the first symmetry breaking. This generates a new quantum number namely 
hyper-isospin. 
\item Since $\susi$ contains $\sufi$ as its sub-group, 
$\tilde{\lambda}_{24}$ should be able to be contained in a $6 \times 6$ 
matrix which is the linear combination of $\lambda_{34}$ and $\lambda_{35}$, 
\ie, 
\be
    c_{34}\lambda_{34} + c_{35}\lambda_{35}
    = \frac{2}{\sqrt{15}} \left(
    \begin{array}{ccccc|c}
           1 &             &             &   &  & 0 \\
                        & 1 &             &   &  & \\
                        &             & 1 &   &  &  \vdots \\
                        &             &             & -\frac{3}{2} &  &  \\
                        &             &        &   & -\frac{3}{2} & 0 \\
     \hline
           0 &  & \cdots &  & 0 & 0 \\
    \end{array} 
    \right) = \left(
    \begin{array}{c|c}
    \tilde{\lambda}_{24} & 0 \\ 
    \hline
    0 & 0 
    \end{array} 
    \right)     \; ,
\ee
where the multiplication factors are choosed to be 
$c_{34} = -1/{\sqrt{5}}$ and $c_{35} = -2/{\sqrt{5}}$.
\item $\lambda_{35}$ represents the hypercharges exist in the strong and 
weak forces with opposite signs. This reflects the property of 
its short and long range interactions. We label this kind of hypercharge as 
$C-$hypercharge.
\item On the other hand, the hypercharge induced by $\lambda_{34}$ 
exists only in the weak sector. We label it as $B-$hypercharge. 
\end{itemize}

\section{Quantum number}
\label{sec:quantum}

Since the $\suthc$ symmetry is kept till the low energy scale, in the sense of 
quantum number there is no new physical consequence on it. 
Then, let us focus on the generators form $\suthh$ relevant for the 
electroweak interaction. We should reconsider the Gell-Mann Okubo relation 
which has been well established within the SM. This relation constitutes 
that the isospin and hypercharge are the constituents of charge, 
\ie $Q = I_3 + \frac{1}{2} Y$. In the present framework we have several 
new hypercharges as mentioned in the preceeding section. This motivates us 
to consider the extended Gell-Mann Okubo relation. 

\begin{figure}[t]
  \centering \includegraphics[width=7.5cm]{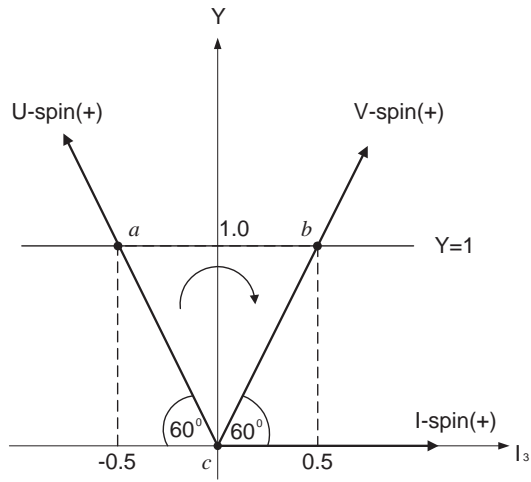}
  \caption{$I-$, $U-$ and $V-$ spin on the $I_3-Y$ plane for $a, b$ and $c$ in the $\suthh$ triplet.}
  \label{fig:spin}
\end{figure}

The $B-$hypercharge induced by $\lambda_{34}$ has non-identical hypercharges, 
\ie $(1 \; \; 1 \; \; -2)$, in contrast with the identical $C-$hypercharge, 
$(1 \; \; 1 \; \; 1)$. If the total hypercharge $Y$ is defined as,
\be
  Y \equiv Y_B + Y_C \; ,
  \label{eq:y}
\ee
we obtain non-identical hypercharge and isospin configurations for the 
doublets which are possibly built from $\suthh$ triplet. This strange 
behaviour can be explained by introducing the $U-$, $V-$ spins beside 
the conventional $I-$spin. If $a,b,c$ denote the first, second and third 
elements in the $\suthh$ triplet, there are three combinations of doublet 
respectively, 
\be
   I-\mathrm{spin} \; : \; 
   \left( \begin{array}{c}
   b \\ 
   a
   \end{array} \right) \; , \; \; 
   U-\mathrm{spin} \; : \; 
   \left( \begin{array}{c}
   a \\ 
   c
   \end{array} \right) \; , \; \; 
   V-\mathrm{spin} \; : \; 
   \left( \begin{array}{c}
   c \\ 
   b
   \end{array} \right) \; .
   \label{eq:spin}
\ee
These combinations can be illustrated on the $I_3-Y$ plane as shown in Fig. 
\ref{fig:spin}. Note that the conventional isospin $I_3$ is determined 
by $\lambda_{29}$. 

From these results, we have found that the third component of hyper-isospin $I_H$ is related with 
isospin and hypercharge as follows, 
\be
 I_{H_3} = I_3 + ( \Delta I_3 \cdot \Delta Y ) \; ,
   \label{eq:ih}
\ee
where the delta means the difference between the upper and lower elements in 
each doublet (see Eq. (\ref{eq:spin})). Secondly, the charge for each element 
can be derived using this hyper-isospin and the total hypercharge, 
\bea
  Q & = & I_{H_3} + \frac{1}{2} Y \\ \nonumber
    & = & I_3 + \left[ \Delta I_3 \cdot \left( \Delta Y_B + \Delta Y_C \right) \right] + 
          \frac{1}{2} \left( Y_B + Y_C \right) \; ,
  \label{eq:ego}
\eea
using Eqs. (\ref{eq:y}) and (\ref{eq:ih}).
This is the extended Gell-Mann Okubo relation in the framework of $\susi$ GUT 
under consideration.

\section{Particle assignment}
\label{sec:particle}

With the extended Gell-Mann Okubo relation at hand, we are now ready 
to go on assigning the particle contents appropriately. As mentioned 
earlier we must fill all fermions in the combination of the 
$\{6\}-$ and $\{\overline{15}\}-$plets. 

Taking into account the quantum numbers (charge, isospin and hypercharge) 
defined above, we should take,  
\be
   (\psi^6)^i_R = \left( 
   \begin{array}{c}
     d^i_r  \\
     d^i_b  \\
     d^i_g  \\
     (\ell^i)^+  \\
    -(\nu^i_\ell)^C \\
     N_{\ell^i}
   \end{array}
   \right)_R \; ,
   \label{eq:6tet}
\ee
for the sextet, while the $\{\overline{15}\}-$plet should consist of,
\be
   (\psi^{15})^{ij}_L = \frac{1}{\sqrt{2}} \left( 
   \begin{array}{cccccc}
          0      & (u^i_g)^C & -(u^i_b)^C & -u^i_r & -d^i_r     & -d^j_r \\
         -(u^i_g)^C & 0     & (u^i_r)^C  & -u^i_b & -d^i_b     & -d^j_b \\
         (u^i_b)^C  & -(u^i_r)^C & 0      & -u^i_g & -d^i_g     & -d^j_g \\
         u^i_r    & u^i_b   & u^i_g    & 0    & (\ell^j)^+    & -(\ell^i)^+ \\
         d^i_r    & d^i_b   & d^i_g    & -(\ell^j)^+ & 0      & (N_{\ell^i})^C \\
         d^j_r    & d^j_b   & d^j_g    &  (\ell^i)^+   & -(N_{\ell^i})^C & 0 
   \end{array}
   \right)_L \; ,
  \label{eq:15plet}
\ee
where $u^i : u, c, t$; $d^i : d, s, b$; $\ell^i : e, \mu, \tau$; 
$N_{\ell^i} : N_e, N_\mu, N_\tau$ and $r, g, b$ denote the colors respectively. 
$N_{\ell}$'s are newly introduced fermions with neutral charges. 
Note that $i,j$ denote the generation and its combination goes cyclic, \ie 
$(i,j) : (1,2) \rightarrow (2,3) \rightarrow (3,1)$. $L$ and $R$ are the 
projection operators, $L \equiv \frac{1}{2} (1 - \gamma_5)$ and 
$R \equiv \frac{1}{2} (1 + \gamma_5)$.

We should make few remarks here. First, we assign the identical 
fermions for two sextets required to avoid the anomaly. 
Secondly, it is clear that this model on its own implies the 
existence of a new neutral fermion, $N_\ell$, to complete its multiplets. 
This exotic fermion 
then could be interpreted as the heavy Majorana neutrino to enable 
the seesaw mechanism naturally. 
Lastly, this is clearly the minimal particle assignment in the 
present model, \ie the minimal $\susi$ GUT. One could also take other 
possibilities by introducing more exotic fermions as done in \cite{su6}.

According to Eqs. (\ref{eq:6tet}) and (\ref{eq:15plet}), 
$(\psi^6)^i_r$ can be written explicitly for each generation as, 
\be
   (\psi^6)^1_R = \left( 
   \begin{array}{c}
     d_r  \\
     d_b  \\
     d_g  \\
     e^+  \\
    -(\nu_e)^C \\
     N_e
   \end{array}
   \right)_R \; ,
   \; \; \; \; \; 
   (\psi^6)^2_R = \left( 
   \begin{array}{c}
     s_r  \\
     s_b  \\
     s_g  \\
     \mu^+  \\
    -(\nu_\mu)^C \\
     N_\mu
   \end{array}
   \right)_R \; ,
   \; \; \; \; \; 
   (\psi^6)^3_R = \left( 
   \begin{array}{c}
     b_r  \\
     b_b  \\
     b_g  \\
     \tau^+  \\
     -(\nu_\tau)^C \\
     N_\tau
   \end{array}
   \right)_R \; ,
\ee
while the contents of $(\psi^{15})^{ij}_L$ are, 
\bea
 (\psi^{15})^{12}_L & = & \frac{1}{\sqrt{2}} \left(
   \begin{array}{cccccc}
          0      & (u_g)^C & -(u_b)^C & -u_r & -d_r     & -s_r \\
         -(u_g)^C & 0     & (u_r)^C  & -u_b & -d_b     & -s_b \\
         (u_b)^C  & -(u_r)^C & 0      & -u_g & -d_g     & -s_g \\
         u_r    & u_b   & u_g    & 0    & \mu^+    & -e^+ \\
         d_r    & d_b   & d_g    & -\mu^+ & 0      & (N_e)^C \\
         s_r    & s_b   & s_g    & e^+   & -(N_e)^C & 0 
   \end{array}
   \right)_L \; , \nonumber \\
 (\psi^{15})^{23}_L & = & \frac{1}{\sqrt{2}} \left(
   \begin{array}{cccccc}
          0      & (c_g)^C & -(c_b)^C & -c_r & -s_r     & -b_r \\
         -(c_g)^C & 0     & (c_r)^C  & -c_b & -s_b      & -b_b \\
         (c_b)^C  & -(c_r)^C & 0      & -c_g & -s_g     & -b_g \\
         c_r    & c_b   & c_g    & 0    & \tau^+    & -\mu^+ \\
         s_r    & s_b   & s_g    & -\tau^+ & 0      & (N_\mu)^C \\
         b_r    & b_b   & b_g    &  \mu^+   & -(N_\mu)^C & 0 
   \end{array}
   \right)_L \; , \\
 (\psi^{15})^{31}_L & = & \frac{1}{\sqrt{2}} \left(
   \begin{array}{cccccc}
          0      & (t_g)^C & -(t_b)^C & -t_r & -b_r     & -d_r \\
         -(t_g)^C & 0      & (t_r)^C  & -t_b & -b_b     & -d_b \\
         (t_b)^C  & -(t_r)^C & 0      & -t_g & -b_g     & -d_g \\
         t_r    & t_b   & t_g    & 0    & e^+    & -\tau^+ \\
         b_r    & b_b   & b_g    & -e^+ & 0      & (N_\tau)^C \\
         d_r    & d_b   & d_g    &  \tau^+   & -(N_\tau)^C & 0 
   \end{array}
   \right)_L \; . \nonumber
\eea

\section{Summary and discussion}
\label{sec:summary}

We have constructed a complete set of generators for the special unitary 
group $\susi$ which is able to unify three forces of our nature. The 
generators have been derived from the first principle of group theory 
by assuming that the symmetry breaking occurs through the intermediate 
stage with $\suthc \otimes \suthh \otimes \uonc$ symmetry. 

Using these generators, we have found the extended Gell-Mann Okubo relation 
which could accomodate all combinations of quantum numbers contained in 
the model including the new hyper-isospin. The relation leads to a unique 
configuration of fermions in the $\{6\}-$ and $\{\overline{15}\}-$plets. 
In order to fill in the multiplets completely we have introduced a new neutral 
fermion for each generation, $N_\ell$, belongs to the triplet in $\suthh$ sub-group. This 
leads to a natural interpretation that $N_\ell$'s are the heavy Majorana 
neutrinos which play an important role to enable the seesaw 
mechanism. 

Further study should be done to obtain a complete lagrangian 
representing the allowed and newly predicted 
interactions in the framework of this GUT model and its phenomenological 
consequences. More detail investigation should also be performed 
to realize two stages of symmetry breakings through for instance 
the Higgs mechanism. These points will be discussed in detail 
in the subsequent paper \cite{andreas1}.

\subsection*{Acknowledgement}

We would like to appreciate A. Mustofa and C. Wijaya for fruitfull discussion 
during the collaboration. This project is partly funded by DIP Kompetitif LIPI 
(fiscal year 2004). 

\newpage

\subsection*{Appendix}

Here, we provide a complete set of matrices which forms generators for 
$\susi$ group. The last $\lambda_{35}$ is written in Eq. (\ref{eq:l35}).

\begin{minipage}[c]{7cm}
  \[
  \lambda_1=  \left( 
  \begin{array}{ccc|ccc}
    0 & 1 & 0 &  &  &  \\
    1 & 0 & 0 &  & (0)_{3\times 3} &  \\
    0 & 0 & 0 &  &  &  \\
    \hline
      &   &   &  &  &  \\
      & (0)_{3\times 3} &  &  & (0)_{3\times 3} &  \\
      &   &   &  &  &  \\
  \end{array}
  \right)
  \] \\
  \[
  \lambda_3=  \left( \begin{array}{ccc|ccc}
    1 & 0 & 0 &  &  &  \\
    0 & -1 & 0 &  & (0)_{3\times 3} &  \\
    0 & 0 & 0 &  &  &  \\\hline
      &   &   &  &  &  \\
      & (0)_{3\times 3} &  &  & (0)_{3\times 3} &  \\
      &   &   &  &  &  \\
  \end{array}\right)
  \] \\
  \[
\lambda_5=  \left( \begin{array}{ccc|ccc}
    0 & 0 & -i &  &  &  \\
    0 & 0 & 0 &  & (0)_{3\times 3} &  \\
    i & 0 & 0 &  &  &  \\\hline
      &   &   &  &  &  \\
      & (0)_{3\times 3} &  &  & (0)_{3\times 3} &  \\
      &   &   &  &  &  \\
  \end{array}\right)
  \] \\
  \[
\lambda_7=  \left( \begin{array}{ccc|ccc}
    0 & 0    & 0  &  &     &  \\
    0 & 0    & -i &  & (0)_{3\times 3} &  \\
    0 & i    & 0  &  &     &  \\\hline
      &      &    &  &     &  \\
      &  (0)_{3\times 3} &    &  & (0)_{3\times 3} &  \\
      &      &    &  &     &  \\
  \end{array}\right)
  \] \\
  \[
\lambda_9=\left(\begin{array}{ccc|ccc}
      &    &   & 1 & 0  & 0 \\
      & (0)_{3\times 3}&   & 0 & 0  & 0 \\
      &    &   & 0 & 0  & 0 \\\hline
    1 & 0  & 0 &   &    &  \\
    0 & 0  & 0 &   & (0)_{3\times 3}&  \\
    0 & 0  & 0 &   &    &  \\
  \end{array}\right)
  \] 
\end{minipage}
  \hspace*{5mm}
\begin{minipage}[c]{7cm}
  \[
  \lambda_2= \left(
  \begin{array}{ccc|ccc}
    0 & -i & 0 &  &  &  \\
    i & 0 & 0 &  & (0)_{3\times 3} &  \\
    0 & 0 & 0 &  &  &  \\
    \hline
      &   &   &  &  &  \\
    & (0)_{3\times 3} &  &  & (0)_{3\times 3} &  \\
      &   &   &  &  &  \\
  \end{array} 
  \right)
  \] \\
  \[
\lambda_4= \left(\begin{array}{ccc|ccc}
    0 & 0 & 1 &  &  &  \\
    0 & 0 & 0 &  & (0)_{3\times 3} &  \\
    1 & 0 & 0 &  &  &  \\\hline
      &   &   &  &  &  \\
      & (0)_{3\times 3} &  &  & (0)_{3\times 3} &  \\
      &   &   &  &  &  \\
\end{array}\right)
  \] \\
  \[
\lambda_6= \left(\begin{array}{ccc|ccc}
    0 & 0 & 0 &  &  &  \\
    0 & 0 & 1 &  & (0)_{3\times 3} &  \\
    0 & 1 & 0 &  &  &  \\\hline
      &   &   &  &  &  \\
      & (0)_{3\times 3} &  &  & (0)_{3\times 3} &  \\
      &   &   &  &  &  \\
\end{array}\right)
  \] \\
  \[
\lambda_8= \frac{1}{\sqrt{3}}\left(\begin{array}{ccc|ccc}
    1 & 0   & 0 &  &     &  \\
    0 & 1   & 0 &  & (0)_{3\times 3} &  \\
    0 & 0   & -2 &  &     &  \\\hline
      &     &   &  &     &  \\
      & (0)_{3\times 3} &   &  & (0)_{3\times 3} &  \\
      &     &   &  &     &  \\
\end{array}\right)
  \] \\
  \[
\lambda_{10}=\left( \begin{array}{ccc|ccc}
      &    &   & -i& 0  & 0 \\
      & (0)_{3\times 3}&   & 0 & 0  & 0 \\
      &    &   & 0 & 0  & 0 \\\hline
    i & 0  & 0 &   &    &  \\
    0 & 0  & 0 &   & (0)_{3\times 3}&  \\
    0 & 0  & 0 &   &    &  \\
  \end{array}\right)
  \]
\end{minipage} 

\newpage

\begin{minipage}[c]{7cm}
  \[
\lambda_{11}=\left(\begin{array}{ccc|ccc}
      &    &   & 0 & 0  & 0 \\
      & (0)_{3\times 3}&   & 1 & 0  & 0 \\
      &    &   & 0 & 0  & 0 \\\hline
    0 & 1  & 0 &   &    &  \\
    0 & 0  & 0 &   & (0)_{3\times 3}&  \\
    0 & 0  & 0 &   &    &  \\
  \end{array}\right)
  \] \\
  \[
\lambda_{13}=\left(\begin{array}{ccc|ccc}
      &    &   & 0 & 0  & 0 \\
      & (0)_{3\times 3}&   & 0 & 0  & 0 \\
      &    &   & 1 & 0  & 0 \\\hline
    0 & 0  & 1 &   &    &  \\
    0 & 0  & 0 &   & (0)_{3\times 3}&  \\
    0 & 0  & 0 &   &    &  \\
  \end{array}\right)
  \] \\
  \[
\lambda_{15}=\left(\begin{array}{ccc|ccc}
      &    &   & 0 & 1  & 0 \\
      & (0)_{3\times 3}&   & 0 & 0  & 0 \\
      &    &   & 0 & 0  & 0 \\\hline
    0 & 0  & 0 &   &    &  \\
    1 & 0  & 0 &   & (0)_{3\times 3}&  \\
    0 & 0  & 0 &   &    &  \\
  \end{array}\right)
  \] \\
  \[
\lambda_{17}=\left(\begin{array}{ccc|ccc}
      &    &   & 0 & 0  & 0 \\
      & (0)_{3\times 3}&   & 0 & 1  & 0 \\
      &    &   & 0 & 0  & 0 \\\hline
    0 & 0  & 0 &   &    &  \\
    0 & 1  & 0 &   & (0)_{3\times 3}&  \\
    0 & 0  & 0 &   &    &  \\
  \end{array}\right)
  \] \\
  \[
\lambda_{19}=\left(\begin{array}{ccc|ccc}
      &    &   & 0 & 0  & 0 \\
      & (0)_{3\times 3}&   & 0 & 0  & 0 \\
      &    &   & 0 & 1  & 0 \\\hline
    0 & 0  & 0 &   &    &  \\
    0 & 0  & 1 &   & (0)_{3\times 3}&  \\
    0 & 0  & 0 &   &    &  \\
  \end{array}\right)
  \] \\
  \[
\lambda_{21}=\left(\begin{array}{ccc|ccc}
      &    &   & 0 & 0  & 1 \\
      & (0)_{3\times 3}&   & 0 & 0  & 0 \\
      &    &   & 0 & 0  & 0 \\\hline
    0 & 0  & 0 &   &    &  \\
    0 & 0  & 0 &   & (0)_{3\times 3}&  \\
    1 & 0  & 0 &   &    &  \\
  \end{array}\right)
  \]
\end{minipage}
\hspace*{5mm}
\begin{minipage}[c]{7cm}
  \[
\lambda_{12}=\left( \begin{array}{ccc|ccc}
      &    &   & 0 & 0  & 0 \\
      & (0)_{3\times 3} &   & -i & 0  & 0 \\
      &    &   & 0 & 0  & 0 \\\hline
    0 & i  & 0 &   &    &  \\
    0 & 0  & 0 &   & (0)_{3\times 3} &  \\
    0 & 0  & 0 &   &    &  \\
  \end{array}\right)
  \] \\
  \[
\lambda_{14}=\left( \begin{array}{ccc|ccc}
      &    &   & 0 & 0  & 0 \\
      & (0)_{3\times 3} &   & 0 & 0  & 0 \\
      &    &   & -i & 0  & 0 \\\hline
    0 & 0  & i &   &    &  \\
    0 & 0  & 0 &   & (0)_{3\times 3} &  \\
    0 & 0  & 0 &   &    &  \\
  \end{array}\right)
  \] \\
  \[
\lambda_{16}=\left( \begin{array}{ccc|ccc}
      &    &   & 0 & -i  & 0 \\
      & (0)_{3\times 3} &   & 0 & 0  & 0 \\
      &    &   & 0 & 0  & 0 \\\hline
    0 & 0  & i &   &    &  \\
    0 & 0  & 0 &   & (0)_{3\times 3} &  \\
    0 & 0  & 0 &   &    &  \\
  \end{array}\right)
  \] \\
  \[
\lambda_{18}=\left( \begin{array}{ccc|ccc}
      &    &   & 0 &  0  & 0 \\
      & (0)_{3\times 3} &   & 0 & -i  & 0 \\
      &    &   & 0 & 0  & 0 \\\hline
    0 & 0  & 0 &   &    &  \\
    0 & i  & 0 &   & (0)_{3\times 3} &  \\
    0 & 0  & 0 &   &    &  \\
  \end{array}\right)
  \] \\
  \[
\lambda_{20}=\left( \begin{array}{ccc|ccc}
      &    &   & 0 &  0  & 0 \\
      & (0)_{3\times 3} &   & 0 &  0 & 0 \\
      &    &   & 0 & -i  & 0 \\\hline
    0 & 0  & 0 &   &    &  \\
    0 & 0  & i &   & (0)_{3\times 3} &  \\
    0 & 0  & 0 &   &    &  \\
  \end{array}\right)
  \] \\
  \[
\lambda_{22}=\left( \begin{array}{ccc|ccc}
      &    &   & 0 &  0  & -i \\
      & (0)_{3\times 3} &   & 0 &  0 & 0 \\
      &    &   & 0 & 0  & 0 \\\hline
    0 & 0  & 0 &   &    &  \\
    0 & 0  & 0 &   & (0)_{3\times 3} &  \\
    i & 0  & 0 &   &    &  \\
  \end{array}\right)
  \]
\end{minipage}

\newpage

\begin{minipage}[c]{7cm}
  \[
\lambda_{23}=\left(\begin{array}{ccc|ccc}
      &    &   & 0 & 0  & 0 \\
      & (0)_{3\times 3} &   & 0 & 0  & 1 \\
      &    &   & 0 & 0  & 0 \\\hline
    0 & 0  & 0 &   &    &  \\
    0 & 0  & 0 &   & (0)_{3\times 3} &  \\
    0 & 1  & 0 &   &    &  \\
  \end{array}\right)
  \] \\
  \[
\lambda_{25}=\left(\begin{array}{ccc|ccc}
      &    &   & 0 & 0  & 0 \\
      & (0)_{3\times 3} &   & 0 & 0  & 0 \\
      &    &   & 0 & 0  & 1 \\\hline
    0 & 0  & 0 &   &    &  \\
    0 & 0  & 0 &   & (0)_{3\times 3} &  \\
    0 & 0  & 1 &   &    &  \\
  \end{array}\right)
  \] \\
  \[
\lambda_{27}=\left(\begin{array}{ccc|ccc}
      &    &   &   &    &  \\
      & (0)_{3\times 3} &   &   &(0)_{3\times 3} &  \\
      &    &   &   &    &  \\\hline
      &    &   & 0 & 1  & 0\\
      &(0)_{3\times 3} &   & 1 & 0  & 0\\
      &    &   & 0 & 0  & 0\\
  \end{array}\right)
  \] \\
  \[
\lambda_{29}=\left(\begin{array}{ccc|ccc}
      &    &   &   &    &  \\
      & (0)_{3\times 3} &   &   &(0)_{3\times 3} &  \\
      &    &   &   &    &  \\\hline
      &    &   & 1 & 0  & 0\\
      &(0)_{3\times 3} &   & 0 & -1  & 0\\
      &    &   & 0 & 0  & 0\\
  \end{array}\right)
  \] \\
  \[
\lambda_{31}=\left(\begin{array}{ccc|ccc}
      &    &   &   &    &  \\
      & (0)_{3\times 3} &   &   &(0)_{3\times 3} &  \\
      &    &   &   &    &  \\\hline
      &    &   & 0 & 0  & -i\\
      &(0)_{3\times 3} &   & 0 & 0  & 0\\
      &    &   & i & 0  & 0\\
  \end{array}\right)
  \] \\
  \[
\lambda_{33}=\left(\begin{array}{ccc|ccc}
      &    &   &   &    &  \\
      & (0)_{3\times 3} &   &   &(0)_{3\times 3} &  \\
      &    &   &   &    &  \\\hline
      &    &   & 0 & 0  & 0\\
      &(0)_{3\times 3} &   & 0 & 0  & -i\\
      &    &   & 0 & i  & 0\\
  \end{array}\right)
  \] 
\end{minipage}
  \hspace*{5mm}
  \hspace*{5mm}
\begin{minipage}[c]{7cm}
  \[
\lambda_{24}=\left( \begin{array}{ccc|ccc}
      &    &   & 0 &  0  & 0 \\
      & (0)_{3\times 3} &   & 0 &  0 & -i \\
      &    &   & 0 & 0  & 0 \\\hline
    0 & 0  & 0 &   &    &  \\
    0 & 0  & 0 &   & (0)_{3\times 3} &  \\
    0 & i  & 0 &   &    &  \\
  \end{array}\right)
  \] \\
  \[
\lambda_{26}=\left( \begin{array}{ccc|ccc}
      &    &   & 0 &  0  & 0 \\
      & (0)_{3\times 3} &   & 0 &  0 & 0 \\
      &    &   & 0 & 0  & -i \\\hline
    0 & 0  & 0 &   &    &  \\
    0 & 0  & 0 &   & (0)_{3\times 3} &  \\
    0 & 0  & i &   &    &  \\
  \end{array}\right)
  \] \\
  \[
\lambda_{28}=\left( \begin{array}{ccc|ccc}
      &    &   &   &    &  \\
      & (0)_{3\times 3} &   &   &(0)_{3\times 3} &  \\
      &    &   &   &    &  \\\hline
      &    &   & 0 & -i  & 0\\
      &(0)_{3\times 3} &   & i & 0  & 0\\
      &    &   & 0 & 0  & 0\\
        \end{array}\right)
  \] \\
  \[
\lambda_{30}=\left( \begin{array}{ccc|ccc}
      &    &   &   &    &  \\
      & (0)_{3\times 3} &   &   &(0)_{3\times 3} &  \\
      &    &   &   &    &  \\\hline
      &    &   & 0 & 0  & 1\\
      &(0)_{3\times 3} &   & 0 & 0  & 0\\
      &    &   & 1 & 0  & 0\\
        \end{array}\right)
  \] \\
  \[
\lambda_{32}=\left( \begin{array}{ccc|ccc}
      &    &   &   &    &  \\
      & (0)_{3\times 3} &   &   &(0)_{3\times 3} &  \\
      &    &   &   &    &  \\\hline
      &    &   & 0 & 0  & 0\\
      &(0)_{3\times 3} &   & 0 & 0  & 1\\
      &    &   & 0 & 1  & 0\\
        \end{array}\right)
  \] \\
  \[
\lambda_{34}=\frac{1}{\sqrt{3}}\left( \begin{array}{ccc|ccc}
      &    &   &   &    &  \\
      & (0)_{3\times 3} &   &   &(0)_{3\times 3} &  \\
      &    &   &   &    &  \\\hline
      &    &   & 1 & 0  & 0\\
      &(0)_{3\times 3} &   & 0 & 1  & 0\\
      &    &   & 0 & 0  & -2\\
        \end{array}\right)
  \]
\end{minipage}


\begin{thebibliography}{99}
	\bibitem{gws} S.L. Glashow, 
		\j{\np}{22}{(1961)}{579}; \\
		S. Weinberg, 
		\j{\prl}{19}{(1967)}{1264}; \\
		A. Salam, 
		\j{Elementary Particle Theory (Ed. N. Svartholm), Almquist and Wiksells, Stockholm}{}{(1969);}{} \\
		S. Glashow, J. Iliopoulos and L. Maiani, 
		\j{\pr}{D2}{(1970)}{1285}.
	\bibitem{pdg} Particle Data Group,  
		\j{Phys. Lett}{B592}{(2004)}{1}.
	\bibitem{massiveneutrino} S. Fukuda \etal (Super-Kamiokande Collaboration), 
		\j{\prl}{81}{(1998)}{1562}; \\
		\textit{ibid.}, 
		\j{\prl}{86}{(2001)}{5651}. 
	\bibitem{nutev} G.P. Zeller \etal (NuTeV Collaboration), 
		\j{\prl}{88}{(2002)}{091802}; \\
                For a comprehensive review on both theoretical and experimental aspects see : G.P. Zeller, 
		\j{PhD Theses, Northwestern University}{}{(2002).}{}
	\bibitem{betadecay} C. Aalseth \etal (Working Group on Neutrinoless double beta decay and direct search for neutrino mass),
		\j{hep-ph/0412300}{}{(2004).}{}
	\bibitem{lhc} V.A. Mitsou  \etal (ATLAS),
		\j{ATLAS TDR on Physics Performance Vol. 2}{}{(1999);}{} \\
		G.L. Bayatian \etal (CMS TP),
		\j{CERN/LHC 94-38}{}{(1994);}{} \\
		E. Accomando \etal, 
		\j{Phys. Rep.}{299}{(1998)}{1}; \\
   		J.A. Aguilar-Saavedra \etal, 
		\j{hep-ph/0106315}{}{(2001);}{} \\
                T. Abe \etal, 
		\j{hep-ph/0109166}{}{(2001);}{} \\
		M. Battaglia, 
		\j{hep-ph/0103338}{}{(2001);}{} \\
 		B. Austin, A. Blondel and J. Ellis (eds.), 
		\j{CERN 99-02}{}{(1999);}{} \\
		C. M. Ankenbrandt \etal, 
		\j{Phys. Rev. ST Accel. Beams}{2}{(1999)}{081001}.
	\bibitem{su5} H. Georgi and S.L. Glashow, 
		\j{\prl}{32}{(1974)}{438}.
	\bibitem{so10} H. Georgi,  
		\j{Particles and Fields (Ed. C.E. Carlson), American Institute of Physics, New York}{}{(1975);}{} \\
		H. Fritzsch and P. Minkowski, 
		\j{\ap}{93}{(1975)}{193}.
	\bibitem{andreas1} A. Hartanto, C. Wijaya and L.T. Handoko, 
		\j{in preparation.}{}{}{}
	\bibitem{su6} M. Fukugita, T. Yanagida and M. Yoshimura, 
		\j{\pl}{B109}{(1982)}{369}.
        \bibitem{su6nn} P. Majumdar,
		\j{\pl}{B121}{(1983)}{25}; \\
		K. Tabata, I. Umemura and K. Yamamoto, 
		\j{Prog. Theor. Phys.}{71}{(1984)}{615}; \\
		R. Barbieri, Z. Berezhiani, G. Dvali, L. Hall and A. Strumia, 
		\j{Nucl.Phys.}{B432}{(1994)}{49}.
	\bibitem{anofree} J. Banks and H. Georgi, 
		\j{\pr}{D14}{(1976)}{1159}; \\
		S. Okubo, 
		\j{\pr}{D16}{(1977)}{3528}.
    	\bibitem{stancu} For example see : F.L. Stancu, 
      		\j{Group Theory in Subnuclear Physics, Oxford Science Publications, Oxford}{}{(1992).}{}
	\bibitem{mohapatra} For example see : R.N. Mohapatra,  
		\j{Unification and Supersymmetry : The Frontiers of Quark-Lepton Physics, Springer Verlag New York Inc.}{}{(1992).}{}
	\bibitem{gellmann} M. Gell-Mann,  
		\j{\pr}{125}{(1962)}{1067}.
 \end{thebibliography}
\end{document}